# INTRODUCTION TO THE POMERON STRUCTURE FUNCTION*

P V Landshoff

DAMTP, University of Cambridge

In many ways the pomeron is like the photon, but there are important differences. Factorisation allows us to define a pomeron structure function, even though the pomeron is not a particle. Although we have a model for the light-quark content of the pomeron, which led to the prediction that a surprisingly large fraction of events at HERA would have an extremely-fast final-state proton, its charm and gluon content will have to be got from experiment. Because the pomeron is not a particle, we cannot derive a momentum sum rule.

*Definition of the pomeron structure function*

This introduction to the pomeron structure function applies to the soft pomeron, whose exchange is responsible for the $s^{0.08}$ rise in total cross-sections[1]. See for example figure 1, which shows the $\gamma p$ cross-section: the curve is the sum of an $s^{0.08}$ term corresponding to the soft pomeron and an $s^{-0.45}$ term corresponding to $(\rho, \omega, f_2, a_2)$ exchange. I do not know to what extent my discussion may apply to any other pomeron, whether it be a less soft one[2][3] or a hard one[4].

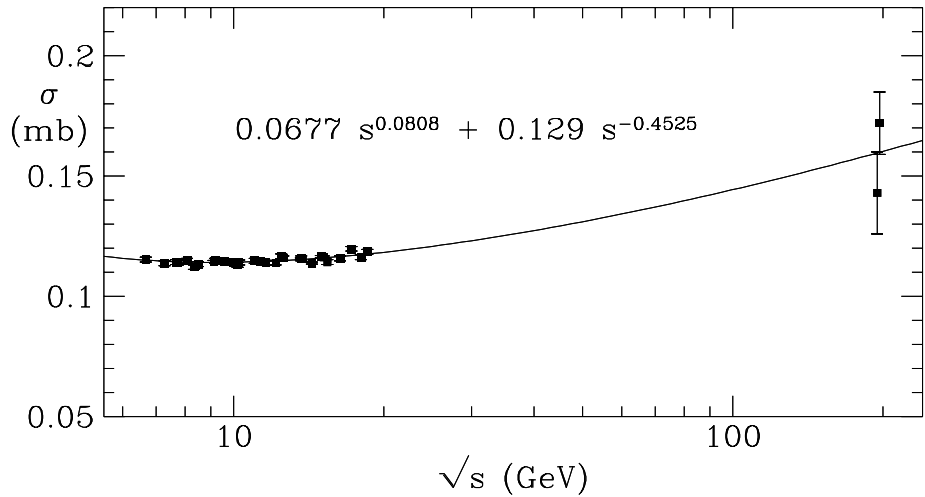

Figure 1: $\gamma p$ total cross-section.

The data for elastic scattering and diffraction dissociation in $pp$ and $\bar{p}p$ collisions are described extremely well[5] by supposing that soft pomeron exchange is similar to photon exchange. For photon exchange between a pair of quarks, the amplitude is

$$\gamma \cdot \gamma \, e^2 \left[\frac{1}{t}\right] \qquad (1a)$$

while for soft pomeron exchange it is

$$\gamma \cdot \gamma \, \beta_0^2 \left[(\alpha' s)^{\alpha(t)-1} \xi_{\alpha(t)}\right] \qquad (1b)$$

Thus the charge $e$ is replaced by a constant pomeron coupling $\beta_0 \approx 2\text{GeV}^{-1}$, and the photon propagator is replaced by the expression in square brackets which is something like a pomeron propagator. Here

$$\alpha(t) = 1 + \epsilon + \alpha' t$$

---

*Based on talks given in April 1995 at Photon '95 (Sheffield) and DIS '95 (Paris)

and $\xi_\alpha$ is a phase factor $-e^{\frac{1}{2}i\pi\alpha}$.

Although photon and pomeron exchange have similarities, there is a crucial difference. The photon propagator has a pole at $t = 0$ because there is a real zero-mass photon, but there is no pole at $t = 0$ in the pomeron propagator because the pomeron is not a particle. We do expect to find a $2^{++}$ particle, probably a glueball, with mass $m$ such that $\alpha(m^2) = 2$, but this particle would have $m \approx 1.9$ GeV and so it has no direct influence on the properties of pomeron exchange near $t = 0$. (However, it is interesting that the WA91 collaboration has reported a $2^{++}$ glueball candidate with just this mass[6][5].)

Consider now the proton structure function $F_2^{\text{proton}}$. Being related to the total $\gamma^* p$ cross-section, it corresponds to a sum over all possible final states. In some small fraction of events, there is an extremely fast proton in the final state[7]. Such events contribute to $F_2^{\text{proton}}(x, Q^2)$ a part which we call $F_2^{\text{diffractive}}(x, Q^2)$. In order to define this we must decide what we mean by an "extremely fast" proton, that is we must specify the maximum fraction $\xi$ of its longitudinal momentum we allow it to lose to include the event. Alternatively, rather than summing over $\xi$ up to some maximaum value, we may introduce $\xi$ as an extra variable into $F_2^{\text{diffractive}}$, and in fact it is useful to introduce also the momentum transfer $t$ between the initial and final protons: $F_2^{\text{diffractive}} = F_2^{\text{diffractive}}(x, Q^2, \xi, t)$.

This definition of $F_2^{\text{diffractive}}$ does not mention the pomeron. By interpreting it in terms of pomeron exchange we find that it has some simple properties: it is leading twist (and so varies only slowly with $Q^2$), and it factorises[8].

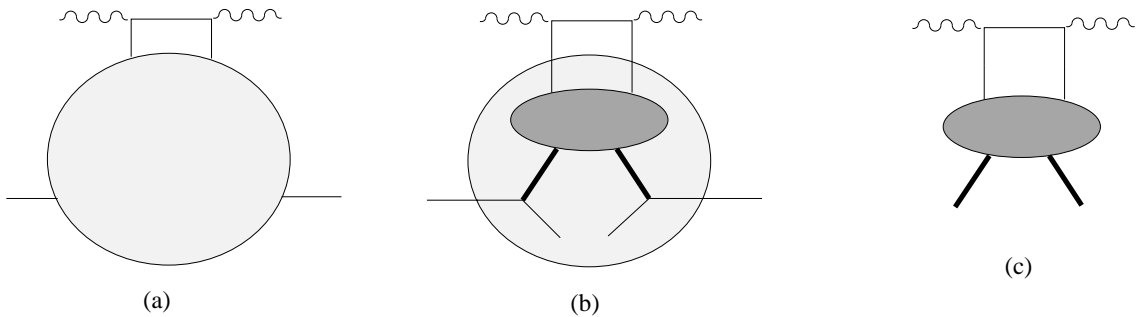

Figure 2: The parton model: (a) $F_2^{\text{proton}}$ (b) $F_2^{\text{diffractive}}$ and (c) $F_2^{\text{pomeron}}$. The black lines represent the pomeron.

In the simple parton model[9], which is a good approximation at not-too-large $Q^2$, $F_2^{\text{proton}}$ corresponds to the diagram in figure 2a. The lower bubble is the amplitude that gives the probability of finding a quark in the proton. It includes all possible nonperturbative contributions. In drawing figure 2a I have used the optical theorem: if we cut the diagram down the middle we reveal the final states. A part of the bubble corresponds to those final states that contain an extremely fast proton, and if we take that part we obtain the diagram of figure 2b. As I have said, this part is leading twist.

The thick lines represent the pomeron. The momentum it carries away from the proton is just $\xi p + \ldots$. In recognition of this it is nowadays usual to rename $\xi$ and instead call it $x_{I\!\!P}$. Soft pomeron exchange has a factorisation property, which yields

$$\frac{d^2}{dt dx_{I\!\!P}} F_2^{\text{diffractive}}(x, Q^2, x_{I\!\!P}, t) = F_{I\!\!P/p}(t, x_{I\!\!P}) F_2^{\text{pomeron}}(\beta, Q^2, t)$$

$$\beta = x/x_{I\!\!P}$$

$$F_{I\!\!P/p}(t, x_{I\!\!P}) = \frac{9\beta_0^2}{4\pi^2}[F_1(t)]^2 x_{I\!\!P}^{1-2\alpha(t)} \tag{3}$$

Here $F_1$ is the Dirac elastic form factor of the proton.

*Momentum sum rule*

We may regard $F_{I\!P/p}(t,x_{I\!P})$ as the flux of pomerons emitted by the proton, and $F_2^{\text{pomeron}}$ as the structure function of the pomeron. In the parton model, $F_2^{\text{pomeron}}$ is just the upper part of figure 2b, drawn in figure 2c. Figure 2c looks just like figure 2a, with the intial-state proton apparently replaced by an intial-state pomeron. However, there is no *particle* called the pomeron; figure 2c rather makes sense because of the factorisation property (3).

In fact, because there is no pomeron particle near $t=0$, this factorisation is not uniquely defined. I think that (3) is the most natural way to define it[8], but we could just as well multiply one of the factors in (3) by any number $N$ and the other by $1/N$. Indeed, the choice $N=\frac{1}{2}\pi$ is found in the literature[10][2]. Because of this ambiguity, one cannot derive a momentum sum rule for $F_2^{\text{pomeron}}$: if such a sum rule were to be satisfied for some choice of $N$, obviously it would not be for any other choice. One cannot derive a momentum sum rule because the pomeron is not a particle.

*Simple model for the pomeron structure function*

I have said that the pomeron in some ways resembles the photon, and so one expects that it has a quark structure function something like that of the photon, in that it consists of two pieces at not-too-large $Q^2$: one that can be calculated from a simple box diagram and another that is not calculable and is most important at small $\beta$.

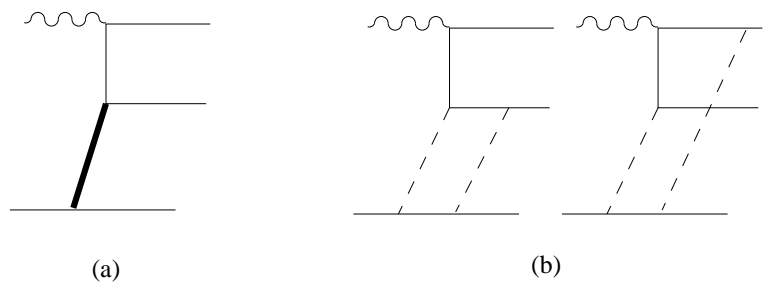

Figure 3: (a) box contribution to $F_2^{\text{pomeron}}$. The black lines represent the pomeron.
(b) the equivalent when pomeron exchange is modelled by two-gluon exchange.

The box contribution corresponds to figure 3a. The simplest model for pomeron exchange is that it corresponds to the exchange of a pair of nonperturbative gluons[11]. Then figure 3a becomes the sum of the two diagrams in figure 3b. Together these give[8][12]

$$\beta q^{\text{pomeron}}(\beta) = C\beta(1-\beta) \quad (5)$$

where $C \approx 0.2$ for each light quark and each light antiquark. This formula provided the remarkable prediction, nearly 10 years ago[8], that some 10% of HERA events would have a very fast final-state proton.

The other piece[8] of the pomeron quark structure function resembles the $\rho$-like contribution to the photon structure function, and so is most important for small $\beta$, where it behaves as $\beta^{-\epsilon}$. If $\beta$ is not *too* small, one expects that $\epsilon=0.08$, as for the proton structure function[13]. But for extremely small $\beta$ one would expect the same to happen as with $F_2^{\text{proton}}$, and see a marked increase in the effective value of $\epsilon$.

*Scaling violation and related issues*

Even though the pomeron is not a particle, the factorisation that leads to figure 2c makes it natural[8][14] to write an evolution equation for $F_2^{\text{pomeron}}$ similar to that for $F_2^{\text{proton}}$, as if the pomeron were the initial state. For this, one needs an input gluon distribution $g^{\text{pomeron}}(\beta)$ at some initial $Q^2$ value. One might guess that its shape is not too different from that of $q^{\text{pomeron}}(\beta)$, but this is just a guess. Further, we have no model that tells us how large it is. Experiment will be important here, for example high-$p_T$ jet production in real-photon diffractive events, the analogue of the UA8 experiment at the CERN $\bar{p}p$ collider[15].

We also need to know what is the initial charmed-quark content of the pomeron. Again we have no reliable model, and so experiment will be important. The coupling of the pomeron to quarks could be rather flavour-blind, so that the $c$-quark content might be quite large even at quite small $Q^2$, but even if the coupling is flavour-blind there may be a suppression because of the mass[12]. We do not know.

This research is supported in part by the EU Programme "Human Capital and Mobility", Network "Physics at High Energy Colliders", contract CHRX-CT93-0357 (DG 12 COMA), and by PPARC.

## References

1  A Donnachie and P V Landshoff, Physics Letters B296 (1992) 227

2  A Capella, A Kaidalov, C Merino and J Tran Thanh Van, HEP-PH 9407372

3  E Gotsman, E M Levin and U Maor, HEP-PH 9503394

4  E A Kuraev, L N Lipatov and V S Fadin, Sov Physics JETP 44 (1976) 443
   L V Gribov, E M Levin and M G Ryskin, Physics Reports 100 (1983) 1

5  P V Landshoff, *The two pomerons*, HEP-PH 9410250

6  WA91 collaboration: S Abatzis et al, Physics Letters B324 (1994) 509

7  G Ingelman and P Schlein, Physics Letters B152 (1985) 256

8  A Donnachie and P V Landshoff, Physics Letters B191 (1987) 309; Nuclear Physics B303 (1988) 634

9  P V Landshoff, J C Polkinghorne and R D Short, Nuclear Physics B28 (1970) 210

10  E L Berger, J C Collins, D Soper and G Sterman, Nuclear Physics B288 (1987) 704

11  P V Landshoff and O Nachtmann, Z Physik C35 (1987) 211

12  N N Nikolaev and B G Zakharov, J Exp Theor Phys 78 (1994) 598 M Diehl, Z Physik C66 (1995) 181

13  A Donnachie and P V Landshoff, Z Physik C61 (1994) 139
    H Abramowicz, E M Levin, A Levy and U Maor, Physics Letters B269 (1991) 465

14  T Gehrmann and W J Stirling, HEP-PH 9503351

15  UA8 collaboration: P Schlein, Nuclear Physics B (Proc Suppl) 33A,B (1993) 41